\documentclass[11pt,onecolumn,twoside]{IEEEtran}
\ifCLASSINFOpdf
   \usepackage[pdftex]{graphicx}
\else
  \usepackage[dvips]{graphicx}
\fi
\usepackage[cmex10]{amsmath}
\usepackage{amssymb}

\newtheorem{proposition}{Proposition}

\begin{document}
%
\title{Bounded Confidence Opinion Dynamics in a \\Social Network of Bayesian Decision Makers}
%
%
%

\author{Kush~R.~Varshney
\thanks{Part of the material in this paper was presented at the Interdisciplinary Workshop on Information and Decision in Social Networks, Cambridge, MA, November 2012 and at the IEEE International Conference on Acoustics, Speech, and Signal Processing, Vancouver, Canada, May 2013.}
\thanks{K.~R.~Varshney is with the Business Analytics and Mathematical Sciences Department, IBM Thomas J.~Watson Research Center, Yorktown Heights, NY, 10598 USA (e-mail: krvarshn@us.ibm.com).}
}

%
%

\markboth{}%
{}
%



\maketitle

\begin{abstract}
Bounded confidence opinion dynamics model the propagation of information in social networks.  However in the existing literature, opinions are only viewed as abstract quantities without semantics rather than as part of a decision-making system.  In this work, opinion dynamics are examined when agents are Bayesian decision makers that perform hypothesis testing or signal detection, and the dynamics are applied to prior probabilities of hypotheses.  Bounded confidence is defined on prior probabilities through Bayes risk error divergence, the appropriate measure between priors in hypothesis testing.  This definition contrasts with the measure used between opinions in standard models: absolute error.  It is shown that the rapid convergence of prior probabilities to a small number of limiting values is similar to that seen in the standard Krause-Hegselmann model.  The most interesting finding in this work is that the number of these limiting values and the time to convergence changes with the signal-to-noise ratio in the detection task.  The number of final values or clusters is maximal at intermediate signal-to-noise ratios, suggesting that the most contentious issues lead to the largest number of factions.  It is at these same intermediate signal-to-noise ratios at which the degradation in detection performance of the aggregate vote of the decision makers is greatest in comparison to the Bayes optimal detection performance.
\end{abstract}


%
\IEEEpeerreviewmaketitle

\section{Introduction}
\label{sec:intro}
%
%
%
%
\IEEEPARstart{P}{eople} are always making choices and decisions on various topics.  These decisions are made based on prior held beliefs, imperfect observations of the world, and utilities associated with correct and incorrect decisions of various types \cite{Ramsey1931,deFinetti1937,vonNeumannM1944,Savage1954}.  Decision making under uncertainty and risk is one of the most basic of human actions.  Mathematically speaking, the decision rule that maximizes expected utility takes the form of a likelihood ratio test \cite{VanTrees1968,BernardoS2000}.  When a Bayesian optimal strategy is employed, the likelihood ratio test involves comparison to a threshold that is a function of prior probabilities.

Likelihood ratio tests are not only optimal detection rules, but psychology experiments suggest that humans do, in fact, use them in their decision making \cite{SwetsTB1961,GlanzerHM2009}.  Psychology experiments also suggest that people are able to use prior probabilities in decision rules when available in natural formats \cite{BraceCT1998}, i.e., they are able to use Bayesian decision rules.  Moreover,  it is suggested in \cite{Viscusi1985} that Bayesian models are useful for analyzing human decision-making behavior.  

Among choices that people are asked to make, the kind that we focus on in this paper is giving opinions:\footnote{In all of these questions, one of the hypotheses is true and the other one is false.} Will \emph{Bhaag Milkha Bhaag} be a blockbuster or a flop?  Who will win the presidential election?  Will proposed bill H.R.2879 pass in Congress? Are vaccinations against measles beneficial for children?  Will peplum jackets be popular this fall?  The job of the opinion giver is to try to report the true hypothesis as his or her opinion.  Thus, such opinion questions can be framed as Bayesian binary hypothesis testing or signal detection problems because human decision makers are tasked with predicting the true hypothesis based on noisy observations of the world and their prior held belief with minimal probability of error or Bayes risk \cite{VanTrees1968}.  In this problem, the likelihood ratio test with a prior probability-based threshold is optimal.  


Thus, an opinion given by a person is a function of both an observation and a prior belief.  Observations can have varying levels of noise.  For some issues and opinions, observable features are very predictive of the hypothesis whereas in others, the observable features are not very predictive at all.  Importantly, people do not always have precise and accurate beliefs about the prior probabilities of hypotheses.  Moreover, these beliefs are influenced by a person's social network \cite{DeGroot1974}.   

One may ask why signal detection questions are of interest rather than more direct consumer choice questions such as whether the decision maker will personally buy a peplum jacket or who the decision maker will vote for.  One reason is that it has been empirically shown that aggregating answers from the first type of question (expectation question) yields more accurate forecasts than the second type of question (intention question) \cite{RothschildW2011}.  Also, the first type of question is more in line with the popular definition of opinion.  In this paper, we study the aggregation of opinions through the theory of distributed Bayesian detection \cite{ChairV1986,ViswanathanV1997,RhimVG2012}.

In society, the prior beliefs of opinion givers are not fixed and independent, but mutable and influenced by their peers.  People sway others one way or the other.  However, most people can only be swayed a little bit at a time; they are not influenced by people with a very different prior.  This non-influence of people with beliefs too far away is known as bounded confidence \cite{HegselmannK2002}.  Thus, we may define a neighborhood of people that influence a person's belief based on the criterion that their beliefs are within some threshold of the person's.  Of course in society at-large, people do indirectly influence others not in their neighborhood via multiple hops in the social network.

Opinion dynamics\hspace{1pt}---\hspace{1pt}how opinions evolve among the people in a social network over time\hspace{1pt}---\hspace{1pt}have been studied in a variety of literatures recently \cite{Lorenz2007}.  In this paper, we focus on a discrete-time model that incorporates the idea of bounded confidence \cite{HegselmannK2002,Krause1997}. Known as the Krause-Hegselmann model, it begins with an initial assignment of opinions for each person; for all people in the society, it works by finding all others whose opinion is within a certain absolute value of one's own and updating the own opinion to be the mean of those opinions.  The model has been analyzed to show that opinions converge in finite time, usually quickly, to a few values \cite{BlondelHT2009}, resulting in clusters of people with the same final opinion.  

The existing literature on bounded confidence opinion dynamics, however, only considers opinions as abstract real-valued numbers without connection to a decision-making formulation.  In this work, it is proposed that opinion-giving be treated as a Bayesian signal detection task and that bounded confidence dynamics be applied to the decision makers' prior probabilities.  Such a view of opinion dynamics through a decision-theoretic lens seems not to exist in the literature \cite{Varshney2013}.

When one is to quantify dissimilarity in prior probabilities in the context of Bayesian hypothesis testing, absolute error is not appropriate because it treats all prior probabilities the same.  The difference between a prior probability of 0.1 and 0.2 leads to a Bayes risk difference that may be profoundly different than the Bayes risk difference between a prior probability of 0.4 and 0.5, which is also different when there are different observation models and signal-to-noise ratios.  Bayes risk error divergence, a member of the family of Bregman divergences, is a criterion by which the dissimilarity of two input prior probabilities is examined appropriately according to detection performance \cite{Varshney2011,VarshneyV2008}.  Thus, in this work, it is proposed that Bayes risk error divergence be used to define bounded confidence.  

Under bounded confidence defined in the sense of Bayes risk error divergence, it is shown in this paper that prior probabilities converge in finite time and quickly to a few limiting values in the same way that opinions converge in the standard Krause-Hegselmann model with bounded confidence defined by absolute error.  Much more interestingly, by defining opinion dynamics in the decision-making setting, it is possible to explore the effect of measurement noise on the convergence.  It is found that the largest number of converged clusters occurs for an intermediate signal-to-noise ratio whereas convergence is to only one value for very low and very high signal-to-noise ratios.  An intermediate signal-to-noise ratio corresponds to an issue that is quite contentious.  Thus, this novel finding in this paper relays the interpretation that more contentious issues lead to more factions or parties.  

We also examine the detection performance of decisions aggregated in the social network through majority vote and find that it is again the contentious issues in which there is the most degradation in detection performance as compared to various qualities of centralized and decentralized Bayesian detection with true priors.  

The remainder of the paper is organized as follows.  First, we provide several preliminaries in Section~\ref{sec:prelim}, including a discussion of relationships to prior work and background material on the Bayes risk error divergence and the Krause-Hegselmann model of bounded confidence opinion dynamics.  In Section~\ref{sec:brebc}, the new model for opinions when viewed as prior probabilities for decision-making tasks is proposed by considering bounded confidence in the Bayes risk error divergence sense.  This section also gives a proof of convergence of the decision weights of decision makers in finite time to clusters and expressions for aggregate decisions.  Simulation results are given in Section~\ref{sec:empirical} that show the dynamics of the proposed model, including the varying number of clusters as a function of signal-to-noise ratio in the Bayesian detection task.  Section~\ref{sec:conclusion} summarizes the contributions and findings, and suggests directions for future work.

\section{Preliminaries}
\label{sec:prelim}

In this section, we provide background on several topics in preparation for proposing the new decision-making opinion dynamics model.  We discuss relationships to prior work in various fields and recapitulate Bayesian detection theory and opinion dynamics models.

\subsection{Relationship to Previous Work}
\label{sec:prelim:previous}

The work presented in this paper relates to a few different areas of previous research.  Within the signal processing, communications, control, and economics fields, there has been much recent mathematical modeling and theoretical research on the spread of ideas in societies ranging from consensus to belief propagation to informational cascade to herding models \cite{AcemogluO2011,ChamleySL2013}.  Bounded confidence opinion dynamics models constitute one class of mathematical models of the spread of beliefs that is currently receiving much attention.  As discussed in Section \ref{sec:intro}, bounded confidence models the fact that people are influenced by others whose opinion is not too different from their own.  Within this class, the Deffuant-Weisbuch model considers pairwise interactions between agents whereas the Krause-Hegselmann model considers interactions among groups of agents all at once \cite{Krause1997,DeffuantNAW2000,HegselmannK2002}; there are both continuous-time and discrete-time versions of the models.  The work herein specifically focuses on the discrete-time Krause-Hegselmann model, but could be extended to the other versions of bounded confidence opinion dynamics models.

There have been several recent extensions of bounded confidence opinion dynamics models including models with prominent agents \cite{AcemogluO2011}, stubborn agents \cite{YildizAOSS2011,FrascaRTI2013},  rebel agents \cite{Eger2013}, agents with vector-valued opinions \cite{VicenteMC2009}, agents with inertia \cite{BarretoMQ2013}, agents with heterogeneous influence \cite{LiangW2012}, and agents that misrepresent their opinions to conform \cite{BuechelHK2013}.  Moreover, there are extensions that introduce models in which only other people with whom a preexisting social connection exists have influence \cite{Weisbuch2004,Rabbat2012}, in which societies have internal hierarchies \cite{Dittmer2001}, in which there are interactions with a few random agents outside the bounded confidence \cite{LiuW2013}, and in which interactions change gradually with confidence rather than at an abrupt bound \cite{SharpanskykhT2013}.  Importantly, all of these extensions of bounded confidence opinion dynamics models do not treat opinions as anything but abstract quantities without semantics.  In this paper, the opinions that evolve dynamically are prior probabilities or decision weights used in setting the thresholds of likelihood ratio tests, and thus have a decision-theoretic meaning.  Any type of meaning given to opinions is wholly absent from the aforecited works.

Bayesian signal detection involving a social network of people who influence each other's priors represents a further point in the progression of the theory that first started with a single detector \cite{VanTrees1968}, then considered distributed or decentralized detection with a handful of nodes such as isolated radar sets \cite{ViswanathanV1997}, and moved on to considering detection via large-scale sensor networks with many cooperative but constrained nodes \cite{ChamberlandV2003}.  In this work, rather than modeling sensor nodes with power and other physical constraints, we model human detectors with their inherent social influences and bounded rationality, i.e., going from decentralized detection in sensor networks to decentralized detection in social networks.  Similarly, work on detection with faulty sensors models physical faults and failures rather than socially-induced ones \cite{HiggerAE2013}.

Although also utilizing the Bayes risk error divergence and investigating incorrect priors used in the threshold of the likelihood ratio test, the model herein can be distinguished from the distributed detection models with quantized priors presented in \cite{RhimVG2012,RhimVG2013}.  First, the problem setup in that body of work is to consider an ensemble of hypothesis testing problems with shared measurement models and error costs, but different true prior probabilities of hypotheses.  An example is a team of referees deciding whether sports players committed fouls; different players have different prior probabilities of committing fouls.  In contrast, the setup here only considers a single hypothesis test, e.g., giving an opinion on the efficacy of inoculation against measles.  In addition, that work is concerned with finding optimal quantizations or categorizations of that ensemble of decision problems for each agent whereas here, we are modeling the spread of beliefs of agents in a social manner.  

In \cite{RhimG2013}, a serial or tandem architecture of decision makers that perform detection based on observing a private observation in addition to the decisions of earlier agents is considered.  The terminal decision maker's detection performance is examined when decision weights that do not match the true prior probability of the hypotheses are used by the agents in the series.  Such a model is in the realm of social learning and herding models \cite{AcemogluDLO2011}, which represent a different class of diffusion of beliefs in social networks than that considered in this work (bounded confidence opinion dynamics).  

\subsection{Bayesian Hypothesis Testing and Bayes Risk Error}
\label{sec:prelim:bht_bre}

Let us consider the standard Bayesian approach to signal detection or hypothesis testing, focusing on the binary hypothesis case \cite{VanTrees1968}.  There is a binary hypothesis $H \in \{h_0,h_1\}$ with prior probabilities $p_0$ and $p_1$, both non-negative such that $p_0 + p_1 = 1$.  The opinion giver or decision maker does not observe the hypothesis directly, but observes a noisy measurement $Y$ that is related to the hypothesis through the likelihood functions $f_{Y|H}(y|H = h_0)$ and $f_{Y|H}(y|H = h_1)$.  The decision maker observes $y$ and tries to determine whether the true hypothesis is $h_0$ or $h_1$ using the decision rule $\hat{h}(y)$.

The cost of deciding $h_1$ when the true hypothesis is $h_0$ (a Type I error) is $c_{10}$ and the cost of deciding $h_0$ when the true hypothesis is $h_1$ (a Type II error) is $c_{01}$.  The cost of correct decisions is zero. The likelihood ratio test decision rule considered is:
\begin{equation}
\label{eq:lrt}
	\frac{f_{Y|H}(y|H = h_1)}{f_{Y|H}(y|H = h_0)} \mathop{\gtreqless}^{\hat{h}(y)=h_1}_{\hat{h}(y)=h_0} \frac{ac_{10}}{(1-a)c_{01}},
\end{equation}
where if the decision weight $a \in [0,1]$ equals $p_0$, the decision rule is Bayes optimal.  People may not get this decision weight correct when giving their opinions.  The resulting Type I error probability of the decision rule, denoted $p_E^{\text{I}}$, is a function of the decision weight $a$ in the threshold of the likelihood ratio test, i.e.~$p_E^{\text{I}}(a)$.  This is similarly the case for the Type II error probability $p_E^{\text{II}}(a)$.  The overall detection error probability weighted by the costs and the true priors, known as the Bayes risk, is:
\begin{equation}
\label{eq:bayesrisk_mismatch}
	J(p_0,a) = c_{10}p_0p_E^{\text{I}}(a) + c_{01}(1-p_0)p_E^{\text{II}}(a).
\end{equation}

The difference between the Bayes risk of the likelihood ratio test with decision weight $a = p_0$ and the Bayes risk of the likelihood ratio test with any general decision weight $a$ is defined to be the Bayes risk error divergence \cite{Varshney2011,VarshneyV2008}:
\begin{equation}
\label{eq:bred}
	d(p_0,a) = J(p_0,a) - J(p_0,p_0).
\end{equation}
The Bayes risk error divergence is the appropriate measure of proximity between prior probabilities of hypotheses because it directly captures degradation in decision-making performance.  It is a Bregman divergence.

\subsection{Krause-Hegselmann Bounded Confidence Model}
\label{sec:prelim:krause}

Let us also consider a now standard approach to modeling opinion dynamics: the Krause-Hegselmann model \cite{HegselmannK2002,Krause1997}.  We have a society of opinion givers $\mathcal{V} = \{1,\ldots,n\}$ with the opinion of person $i$ at time $t$ being $a_i(t) \in \mathbb{R}$. Each opinion giver has a neighborhood of other opinion givers in the society whose opinion does not differ by more than an opinion threshold $\theta$:
\begin{equation}
\label{eq:neighborhood}
	\mathcal{N}_i(t) = \{j \in \mathcal{V} \mid |a_i(t) - a_j(t)| \le \theta\}.
\end{equation}
The restriction by the threshold $\theta$ is the model for bounded confidence.

Each opinion giver begins with an initial opinion.  The opinions are updated based on interaction and influence from neighbors.  The opinion giver's opinion is updated at the next time step to the mean or centroid of the neighborhood at the current time step:
\begin{equation}
\label{eq:krause_update}
	a_i(t+1) = \frac{1}{|\mathcal{N}_i(t)|} \sum_{j\in \mathcal{N}_i(t)} a_j(t).
\end{equation}
It has been shown that these opinions converge in a finite number of steps to a few clusters with all opinion givers in each cluster sharing the same opinion value \cite{BlondelHT2009}.

\section{Bounded Confidence in the Sense of Bayes Risk Error Divergence}
\label{sec:brebc}

We now propose a bounded confidence opinion dynamics model for Bayesian decision makers by endowing opinion-giving with a decision-making task, specifically detection.  We do so by combining the ideas discussed in Section \ref{sec:prelim:bht_bre} and Section~\ref{sec:prelim:krause}, specifically by having opinion givers observe a noisy measurement $Y$ and use the likelihood ratio test decision rule \eqref{eq:lrt}, and have their personal prior beliefs evolve according to Krause-Hegselmann dynamics.  

\subsection{Proposed Model}
\label{sec:brebc:model}

The key difference in the dynamics from the standard Krause-Hegselmann model is in the quantification of proximity for bounded confidence.  When opinions have a semantic meaning as prior probabilities in a detection task, absolute error as in \eqref{eq:neighborhood} is not how similarity between opinions should be measured.  As discussed earlier, the divergence between prior probabilities that takes detection performance into account is the Bayes risk error.  Working with absolute error and Bayes risk error can lead to quite different outcomes, evidenced in \cite{VarshneyV2008}.  Thus, we take the bounded confidence neighborhood to be defined by Bayes risk error divergence:
\begin{equation}
\label{eq:neighborhood_BRE}
	\mathcal{N}_i(t) = \{j \in \mathcal{V} \mid d(a_i(t),a_j(t)) \le \theta\}.
\end{equation}
It should be noted that due to the likelihood functions in the decision rule and consequently Bayes risk, the Bayes risk error divergence changes with the magnitude of observation noise.

The second part of the Krause-Hegselmann dynamics is the opinion update to the centroid of the neighborhood.  The Bayes risk error divergence centroid of a set of prior probabilities is simply the mean value due to a property of the family of Bregman divergences \cite{BanerjeeMDG2005}, of which the Bayes risk error divergence is a member.  Therefore, the opinion update \eqref{eq:krause_update} remains unchanged from the standard Krause-Hegselmann model when Bayes risk error is considered.  Other means or centroids could be considered, but are beyond the scope of this work \cite{HegselmannK2005,VarshneyV2013}.

\subsection{Convergence Results}
\label{sec:brebc:convergence}

In this section, we present basic convergence properties of the opinion dynamics system with bounded confidence defined in the Bayes risk error divergence sense.  The first proposition states that the initial ordering of the decision weights is preserved throughout the dynamic evolution.  The second proposition states that the dynamics of the decision weights converge in finite time to a few clusters.  The propositions and proof techniques are almost the same as those for the standard Krause-Hegselmann model presented in \cite{BlondelHT2009}.  

\begin{proposition}
\label{prop:order}
	Let $\{a_1(t),\ldots,a_n(t)\}$ evolve according to the dynamics  \eqref{eq:krause_update}--\eqref{eq:neighborhood_BRE}.  If $a_i(0) \le a_j(0)$, then $a_i(t) \le a_j(t)$ for all $t$.
\end{proposition}
\begin{IEEEproof}
	The proof is by induction.  The base case is in the theorem statement: $a_i(0) \le a_j(0)$.  We assume that $a_i(\tau) \le a_j(\tau)$ and show that $a_i(\tau + 1) \le a_j(\tau + 1)$ must then also be true.

Let 
\begin{align*}
	\mathcal{S}_i(\tau) &= \mathcal{N}_i(\tau) \setminus \mathcal{N}_j(\tau), \\
	\mathcal{S}_j(\tau) &= \mathcal{N}_j(\tau) \setminus \mathcal{N}_i(\tau), \\
	\mathcal{S}_{ij}(\tau) &= \mathcal{N}_i(\tau) \cap \mathcal{N}_j(\tau),
\end{align*}
and also let
\begin{align*}
	\bar{a}_{\mathcal{S}_i}(\tau) &= \frac{1}{|\mathcal{S}_i(\tau)|} \sum_{\ell\in \mathcal{S}_i(\tau)} a_{\ell}(\tau), \\
	\bar{a}_{\mathcal{S}_j}(\tau) &= \frac{1}{|\mathcal{S}_j(\tau)|} \sum_{\ell\in \mathcal{S}_j(\tau)} a_{\ell}(\tau), \\
	\bar{a}_{\mathcal{S}_{ij}}(\tau) &= \frac{1}{|\mathcal{S}_{ij}(\tau)|} \sum_{\ell\in \mathcal{S}_{ij}(\tau)} a_{\ell}(\tau).
\end{align*}

As a Bregman divergence, although not symmetric and not convex in the second argument like absolute error, Bayes risk error divergence is monotonically non-decreasing in the second argument.  This monotonicity leads to the conclusion with neighborhoods defined by Bayes risk error divergence that for any $\ell_1 \in \mathcal{S}_i(\tau)$, $\ell_2 \in \mathcal{S}_{ij}(\tau)$, and $\ell_3 \in \mathcal{S}_j(\tau)$, $a_{\ell_1}(\tau) \le a_{\ell_2}(\tau) \le a_{\ell_3}(\tau)$.  Therefore, it is also true that $\bar{a}_{\mathcal{S}_i}(\tau) \le \bar{a}_{\mathcal{S}_{ij}}(\tau) \le \bar{a}_{\mathcal{S}_j}(\tau)$. 

Carrying out the update \eqref{eq:krause_update}, we find that
\begin{align*}
	a_i(\tau+1) &= \frac{|\mathcal{S}_i(\tau)|\bar{a}_{\mathcal{S}_i}(\tau) + |\mathcal{S}_{ij}(\tau)|\bar{a}_{\mathcal{S}_{ij}}(\tau)}{|\mathcal{N}_i(\tau)|} \le \bar{a}_{\mathcal{S}_{ij}}(\tau), \\
	a_j(\tau+1) &= \frac{|\mathcal{S}_j(\tau)|\bar{a}_{\mathcal{S}_j}(\tau) + |\mathcal{S}_{ij}(\tau)|\bar{a}_{\mathcal{S}_{ij}}(\tau)}{|\mathcal{N}_j(\tau)|} \ge \bar{a}_{\mathcal{S}_{ij}}(\tau).
\end{align*}
Therefore, since both right sides are the same, $a_i(\tau+1) \le a_j(\tau+1)$.
\end{IEEEproof}

\begin{proposition}
\label{prop:convergence}
	Let $a_1(t) \le \cdots \le a_n(t)$ evolve according to the dynamics \eqref{eq:krause_update}--\eqref{eq:neighborhood_BRE}.  For all $i$, $a_i(t)$ converges to a limit $a_i^*$ in finite time.  Additionally, for any pair $i$ and $j$, either $a_i^* = a_j^*$ or $\min\{d(a_i^*,a_j^*),d(a_j^*,a_i^*)\} \ge \theta$.
\end{proposition}
\begin{IEEEproof}
	First of all, due to Proposition \ref{prop:order}, $a_1(t)$ is nondecreasing and is bounded from above by $a_n(t)$.  Therefore, it converges to some limiting value $a_1^*$.

	If $\nu \neq n$ is the largest index value of decision makers for which $a_{\nu}$ converges to $a_1^*$, then it is claimed that there is a time such that $\min\{d(a_{\nu},a_{\nu+1}),d(a_{\nu+1},a_{\nu})\} \ge \theta$.

	To prove the claim by contradiction, suppose that $\min\{d(a_{\nu},a_{\nu+1}),d(a_{\nu+1},a_{\nu})\} < \theta$.  Then there is some time after which $\min\{d(a_{i},a_1^*),d(a_1^*,a_{i})\}$ is less than $\epsilon$ for $i = 1,\ldots,\nu$, where $\epsilon$ is a fixed constant greater than zero.  Using the supposition, the decision maker $\nu+1$ is within the bounded confidence of decision makers $1,\ldots,\nu$, and thus at the next time step $\tau+1$, 
\begin{displaymath}
	a_{\nu}(\tau+1) \ge \frac{1}{\nu+1}\sum_{j=1}^{\nu+1}a_j(\tau).
\end{displaymath}
Moreover, because $a_{\nu+1}$ does not converge to $a_1^*$, there is some time $\tau$ at which  $\min\{d(a_{\nu+1},a_1^*),d(a_1^*,a_{\nu+1})\} > \delta > 0$.  Let $a_{\epsilon}$ be the decision weight less than $a_1^*$ such that $\min\{d(a_{\epsilon},a_1^*),d(a_1^*,a_{\epsilon})\} = \epsilon$ and let $a_{\delta}$ be the decision weight greater than $a_1^*$ such that $\min\{d(a_{\delta},a_1^*),d(a_1^*,a_{\delta})\} = \delta$.  Then:
\begin{displaymath}
	\frac{1}{\nu+1}\sum_{j=1}^{\nu+1}a_j(\tau) \ge \frac{1}{\nu+1}\left(a_{\delta} + \sum_{j=1}^{\nu}a_{\epsilon}\right) = \frac{a_{\delta} + \nu a_{\epsilon}}{\nu + 1}.
\end{displaymath}
If we take $\epsilon$ small enough, then $(a_{\delta} + \nu a_{\epsilon})/(\nu + 1) > a_{\epsilon}$.  

Thus overall, under the supposition, we have that $a_{\nu}(\tau + 1) > a_{\epsilon}$, which is a contradiction because we require $a_{\nu}$ to remain such that $\min\{d(a_{\nu},a_1^*),d(a_1^*,a_{\nu})\} < \epsilon$.  Therefore, there is a time such that $\min\{d(a_{\nu},a_{\nu+1}),d(a_{\nu+1},a_{\nu})\} \ge \theta$.  After that time, $a_{\nu}$ cannot increase and $a_{\nu+1}$ cannot decrease, so $\min\{d(a_{\nu},a_{\nu+1}),d(a_{\nu+1},a_{\nu})\} \ge \theta$ for all subsequent times.

Also after that time, the dynamics of decision makers $\nu+1,\ldots,n$ are independent of the dynamics of decision makers $1,\dots,\nu$, so the same argument can be applied to that set of decision makers, and further recursively.  The convergence time is finite because once a set of decision makers is no longer interacting with any decision maker outside of the set, all members obtain the same decision weight on the next time step.
\end{IEEEproof}

\subsection{Aggregate Bayes Risk}
\label{sec:brebc:aggregate}

Each decision maker gives an opinion on the topic one way or the other as his or her local decision, but we can aggregate those responses by majority vote in order to obtain a global decision \cite{RothschildW2011}.  Such aggregation through majority vote falls under the purview of decentralized detection and data fusion \cite{ViswanathanV1997}.  In this section, we give the expression for the Bayes risk of the majority vote of the $n$ people after their decision weights have converged to a small number of clusters.

First, let us explicitly state the majority vote global detection rule that takes local detections as input.  Let us denote the local decisions as $\hat{h}_j(y_j)$, $j=1,\ldots,n$, which are all made according to the likelihood ratio test \eqref{eq:lrt}.  Known as a fusion rule, the majority vote is:
\begin{equation}
\label{eq:fusionrule}
	\hat{h} = \begin{cases} 1\, & \sum_{j=1}^n \hat{h}_j \ge \tfrac{n}{2} \\
			0, & \sum_{j=1}^n \hat{h}_j < \tfrac{n}{2}
		\end{cases}.
\end{equation}
Let the number of converged clusters from the opinion dynamics be $m$, with the number of decision makers in each cluster $n_1^*,\ldots,n_m^*$.  Also let the Type I error of decision makers in cluster $i$ be $p_i^{\text{I}}$ and the Type II error be $p_i^{\text{II}}$.

The global Type I error probability is the sum of the Type I error probabilities for all instances of the number of $\hat{h}_j$ being greater than or equal to $n/2$.  In each of these instances, the Type I error probability is the convolution of the Type I error probabilities of each cluster, which takes the form of a binomial distribution with parameters equal to the number of decision makers in the cluster and the Type I error probability of the decision makers in the cluster.  For the case of one final cluster (where $n_1^* = n$):
\begin{equation}
\label{eq:oneclusterTypeI}
	p_E^{\text{I}} = \sum_{j = \lceil n/2\rceil}^n \binom{n_1^*}{j}{p_1^{\text{I}}}^j\left(1 - p_1^{\text{I}}\right)^{n_1^*-j},
\end{equation}
for the case of two final clusters:
\begin{equation}
\label{eq:twoclustersTypeI}
	p_E^{\text{I}} = \sum_{j = \lceil n/2\rceil}^n \sum_{k=0}^{j} \binom{n_1^*}{j-k}{p_1^{\text{I}}}^{(j-k)}\left(1 - p_1^{\text{I}}\right)^{n_1^*-(j-k)}\binom{n_2^*}{k}{p_2^{\text{I}}}^{k}\left(1 - p_2^{\text{I}}\right)^{n_2^*-k},
\end{equation}
for the case of three final clusters:
\begin{multline}
\label{eq:threeclustersTypeI}
	p_E^{\text{I}} = \\\sum_{j = \lceil n/2\rceil}^n \sum_{k=0}^{j} \binom{n_1^*}{j-k}{p_1^{\text{I}}}^{(j-k)}\left(1 - p_1^{\text{I}}\right)^{n_1^*-(j-k)}\sum_{l=0}^k \binom{n_2^*}{k-l}{p_2^{\text{I}}}^{(k-l)}\left(1 - p_2^{\text{I}}\right)^{n_2^*-(k-l)}\binom{n_3^*}{l}{p_3^{\text{I}}}^{l}\left(1 - p_3^{\text{I}}\right)^{n_3^*-l},
\end{multline}
and so on.  The same expressions apply for the Type II error with every instance of `I' replaced by `II.'  The two error probabilities are then combined, weighted by the costs and true priors as in \eqref{eq:bayesrisk_mismatch}, to obtain the final aggregate Bayes risk with converged clusters.

For comparison, we can compare this aggregate Bayes risk with converged clusters to various optimal centralized and decentralized Bayes risks using true priors.  The centralized Bayes risk is obtained by simply calculating error probabilities with a combined $n$-dimensional measurement vector with independent components.  In Section~\ref{sec:empirical}, we examine the difference between aggregate Bayes risk with converged clusters and the centralized Bayes risk; this difference is another kind of Bayes risk error for distributed detection.  

The decentralized Bayes risk with a majority vote fusion rule has the same expression as the single converged cluster case \eqref{eq:oneclusterTypeI}, but with the Bayes optimal error probabilities inserted for $p^{\text{I}}$.  One other form of decentralized detection is to change the fusion rule to be optimal under the restriction that the local decisions are the identical local Bayes optimal ones \cite{ChairV1986}.  In this form, known as the Chair-Varshney rule, instead of the majority vote \eqref{eq:fusionrule}, the fusion rule is:
\begin{equation}
\label{eq:fusionrule}
	\hat{h} = \begin{cases} 1\, & \sum_{j=1}^n \hat{h}_j \ge \gamma \\
			0, & \sum_{j=1}^n \hat{h}_j < \gamma
		\end{cases},
\end{equation}
where
\begin{displaymath}
	\gamma = \frac{\log\left(\frac{p_0c_{10}}{(1-p_0)c_{01}}\right) - n\log\left(\frac{p^{\text{II}}}{1 - p^{\text{I}}}\right)}{\log\left(\frac{p^{\text{I}}p^{\text{II}} - p^{\text{I}} - p^{\text{II}} + 1}{p^{\text{I}}p^{\text{II}}}\right)}.
\end{displaymath}
The Bayes risk of this rule takes the same form as for the Bayes optimal majority vote and the aggregate converged single cluster majority vote \eqref{eq:oneclusterTypeI}, but with the outer sum starting at $j = \lceil\gamma\rceil$.

\section{Empirical Results}
\label{sec:empirical}

In this section, we present simulated examples to illustrate the proposed opinion dynamics model with bounded confidence defined via Bayes risk error divergence.  We consider a set of $n = 101$ opinion givers with initial prior probabilities sampled over the interval $[0,1]$.  We show two examples; we consider the uniform distribution of initial decision weights in the first example and the beta distribution with parameters $2/3$ and $1$ in the second example.  The signal detection task is as follows.  Conditioned on hypothesis $h_0$, $y$ is drawn from a scalar Gaussian distribution with mean $0$ and standard deviation $\sigma$.  Conditioned on hypothesis $h_1$, $y$ is also Gaussian with standard deviation $\sigma$, but with mean $1$.  All decision makers have independent and identically distributed measurements.  All decision makers also have the same error costs $c_{10}$ and $c_{01}$, which are taken to be equal.  The true prior $p_0 = 1/2$ in the first example and $p_0 = 2/5$ in the second example; the true prior matches the expected value of the initial decision weights of the decision makers in both examples.

\begin{figure}
	\begin{center}
		\includegraphics[width=0.47\textwidth]{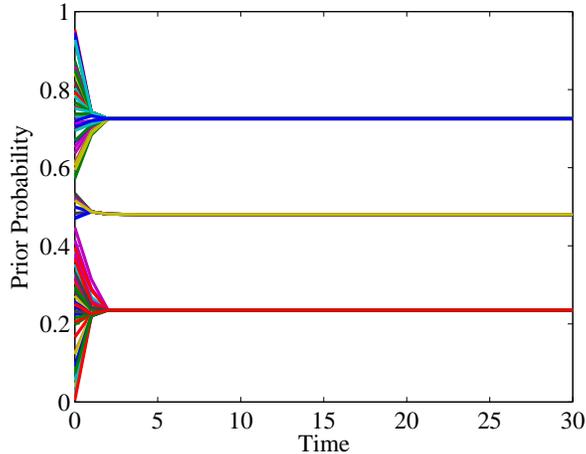}
	\end{center}
	\caption{Dynamics of 101 people with bounded confidence threshold 0.1 for Bayes risk error divergence Krause-Hegselmann model with univariate Gaussian observations having means 0 and 1, and standard deviation 4.}
	\label{fig:dyn1}
\end{figure}

Fig.~\ref{fig:dyn1} shows the dynamics with neighborhoods defined by Bayes risk error as proposed in this paper with a particular value for the observation noise standard deviation $\sigma = 4$. 
 The opinion threshold is fixed at $\theta = 0.1$.  The convergence behavior is qualitatively similar to the standard Krause-Hegselmann model with bounded confidence defined using absolute error.  There is a quick convergence to a few clusters while maintaining the initial ordering under Bayes risk error bounded confidence as well.

\begin{figure}
	\begin{center}
		\begin{tabular}{c}
			\includegraphics[width=0.47\textwidth]{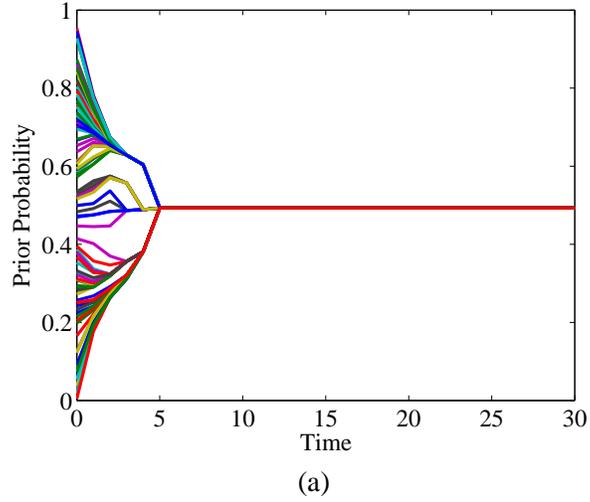} \\
			(a) \\
			\includegraphics[width=0.47\textwidth]{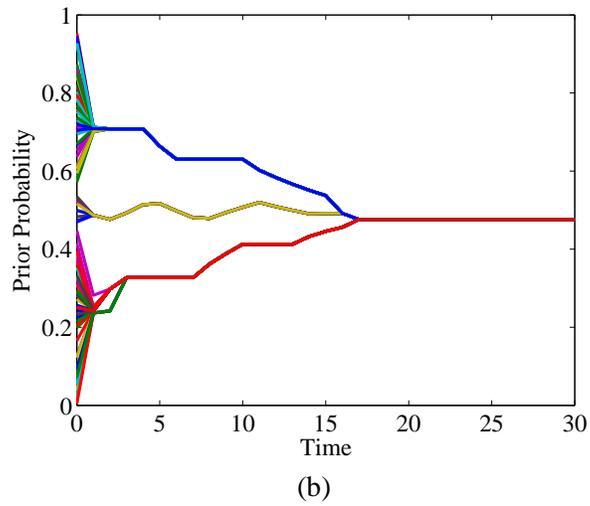} \\
			(b)
		\end{tabular}
	\end{center}
	\caption{Dynamics of 101 people with bounded confidence threshold 0.1 for Bayes risk error divergence Krause-Hegselmann model with univariate Gaussian observations having means 0 and 1, and (a) standard deviation 1 and (b) standard deviation 16.}
	\label{fig:dyn2}
\end{figure}

\begin{figure}
	\begin{center}
		\includegraphics[width=0.47\textwidth]{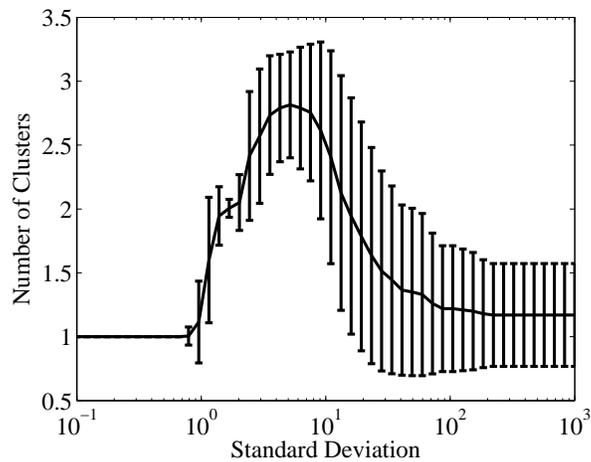}
	\end{center}
	\caption{Converged number of clusters as a function of the noise standard deviation in the first example, averaged over 200 trials.  The error bars indicate the standard deviation of the number of clusters over the trials.}
	\label{fig:numclusters}
\end{figure}

Since (unlike absolute error) the Bayes risk error divergence depends on the likelihood functions of the observation model, bounded confidence opinion dynamics can be investigated under different signal-to-noise ratios.  For smaller and greater noise levels, as seen in Fig.~\ref{fig:dyn2}, the convergence pattern as well as the number of final clusters is different than an intermediate amount of noise as in Fig.~\ref{fig:dyn1}.  
The number of converged clusters (averaged across several trials of initial decision weights) as a function of noise level for a fixed bounded confidence threshold is shown in Fig.~\ref{fig:numclusters}.  
The number of clusters first increases and then decreases with the noise level.  When observations are either very certain or very uncertain, there is convergence to a single cluster; however, in the intermediate case there are more clusters.  

\begin{figure}
	\begin{center}
		\includegraphics[width=0.47\textwidth]{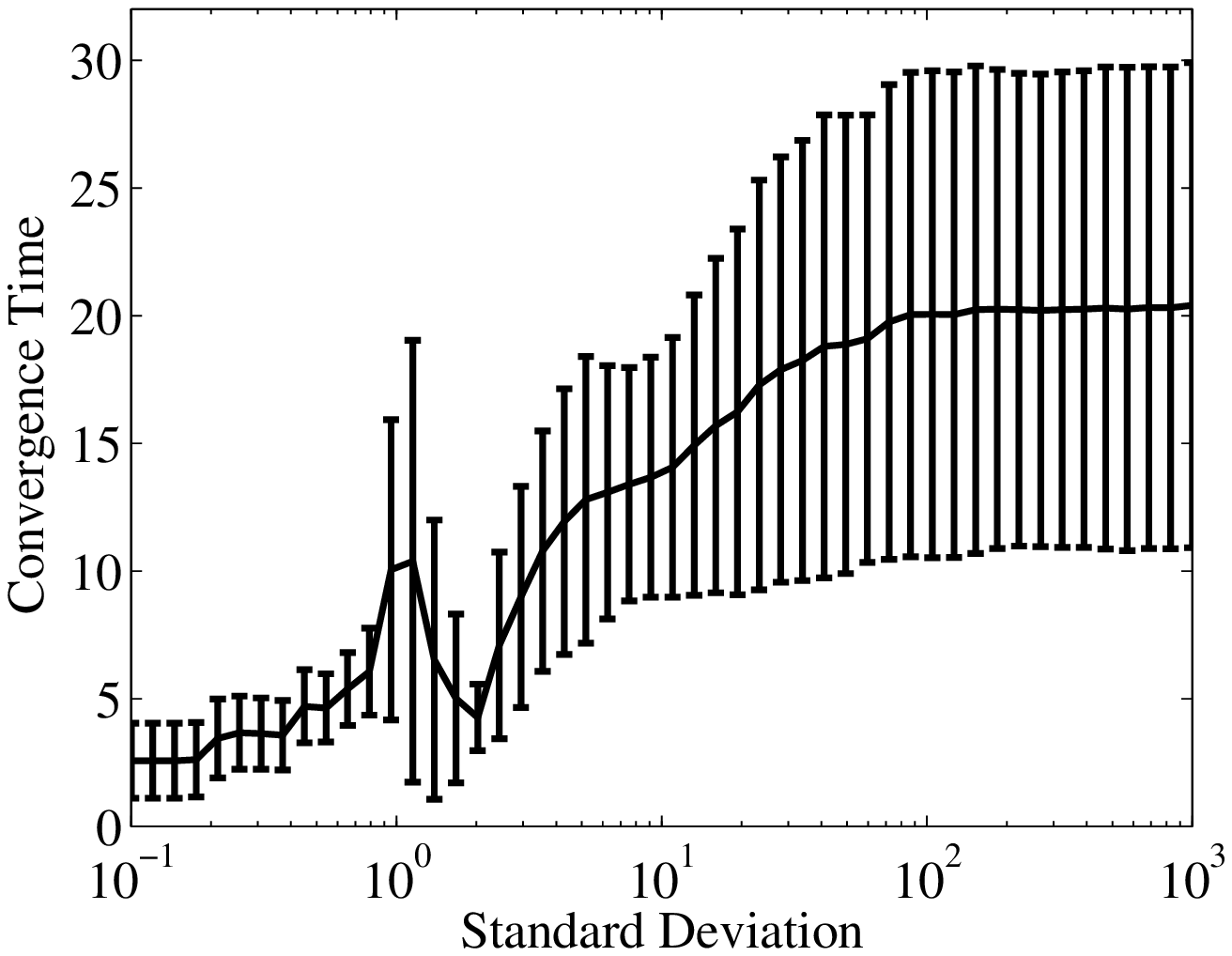}
	\end{center}
	\caption{Convergence time as a function of the noise standard deviation in the first example, averaged over 200 trials.  The error bars indicate the standard deviation of the number of clusters over the trials.}
	\label{fig:numiters}
\end{figure}

\begin{figure}
	\begin{center}
		\includegraphics[width=0.47\textwidth]{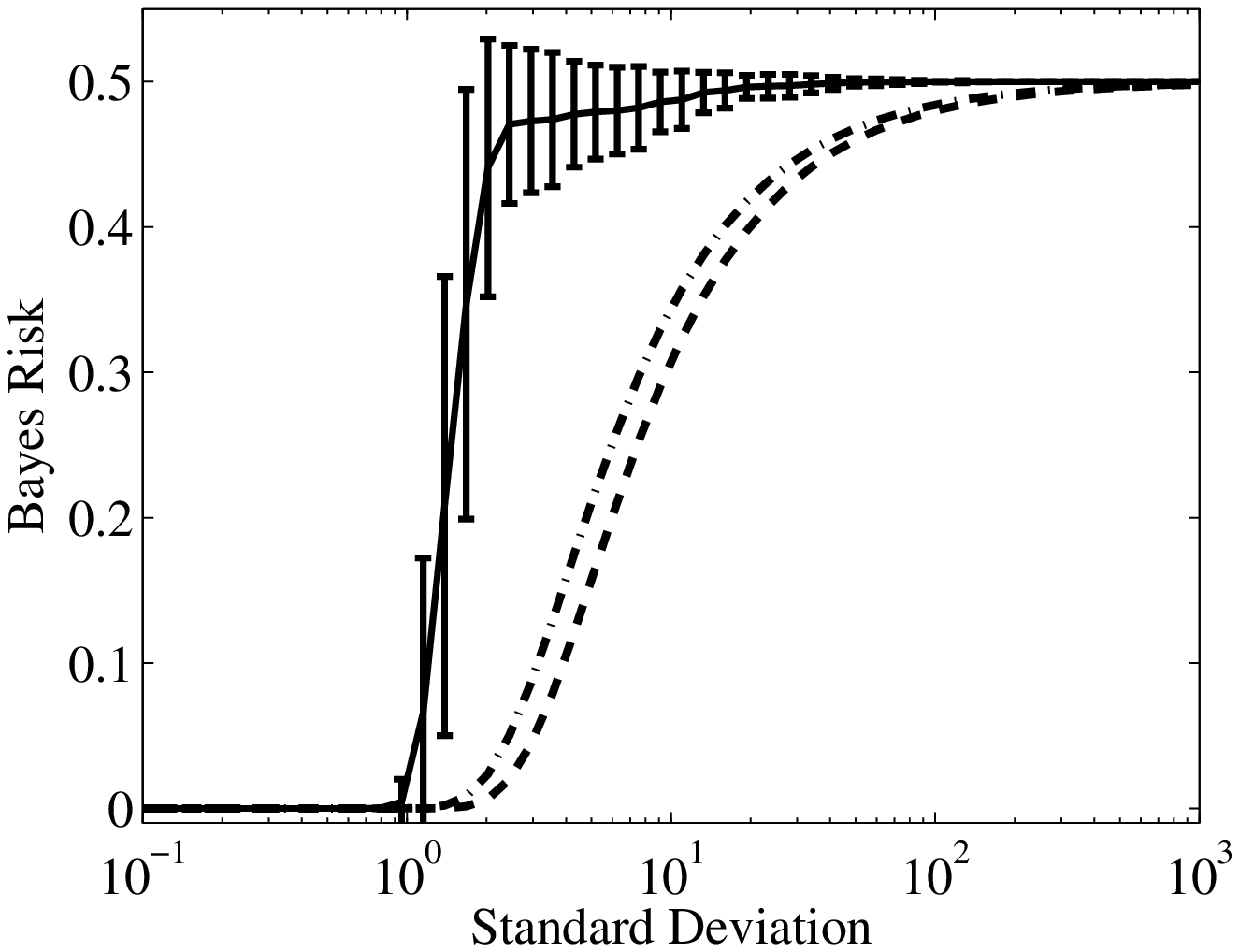}
	\end{center}
	\caption{Bayes risk at convergence of the majority vote of 101 opinion givers as a function of the noise standard deviation in the first example, averaged over 200 trials.  The error bars indicate the standard deviation of the number of clusters over the trials.  The dashed line is the centralized Bayes optimal risk and the dashed and dotted line is the decentralized majority vote Bayes optimal risk.}
	\label{fig:bayesrisk}
\end{figure}

\begin{figure}
	\begin{center}
		\includegraphics[width=0.47\textwidth]{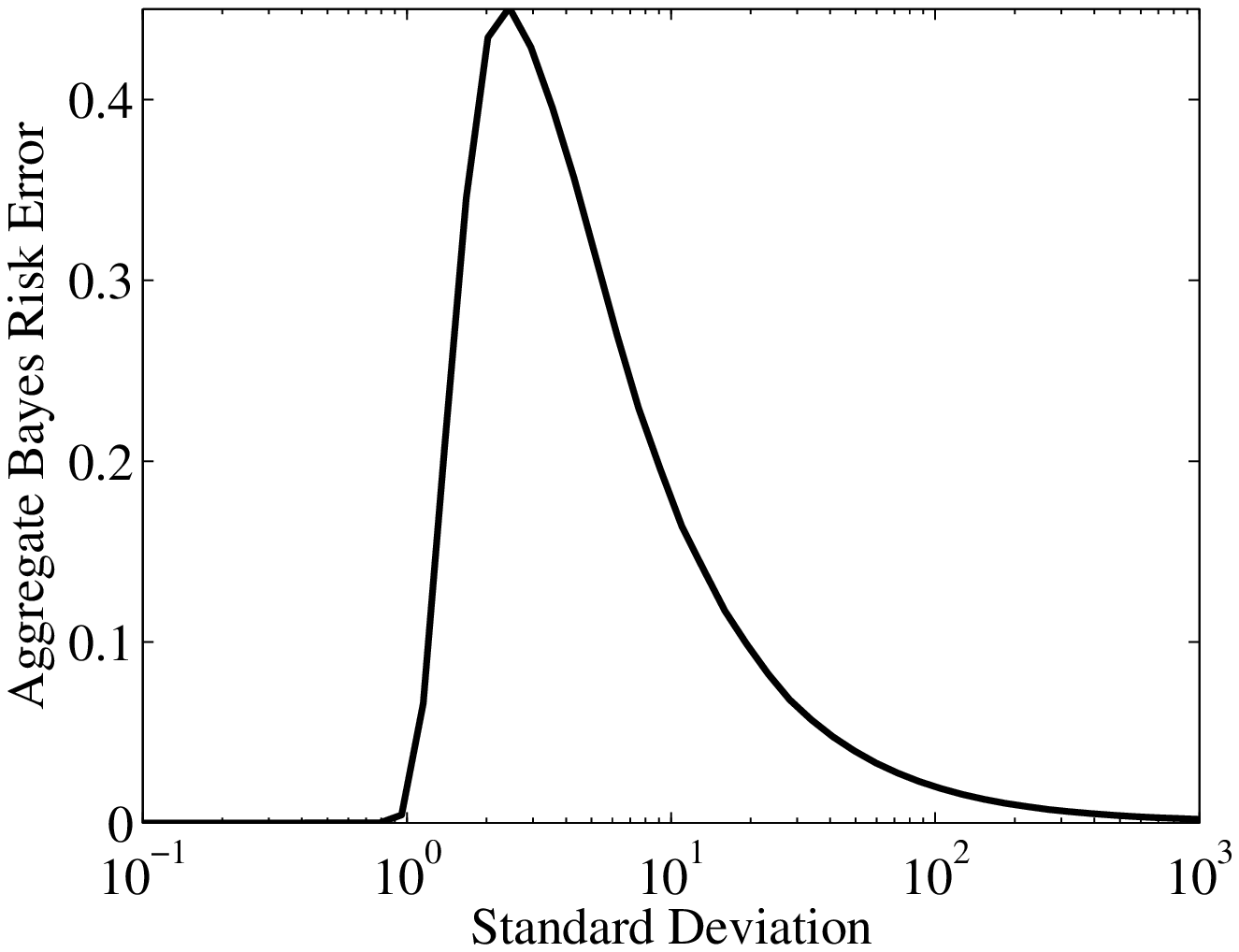}
	\end{center}
	\caption{Aggregate Bayes risk error at convergence of the majority vote of 101 opinion givers as a function of the noise standard deviation in the first example, averaged over 200 trials.}
	\label{fig:aggbre}
\end{figure}

The number of time steps to convergence is also a function of the signal-to-noise ratio, as seen in Fig.~\ref{fig:numiters}.
The Bayes risk of the aggregate majority vote of the opinion givers is shown in Fig.~\ref{fig:bayesrisk} once the clusters have converged.
For comparison, the Bayes risks of the centralized detection rule and of the decentralized majority vote with true priors used in the local detection thresholds are also shown.  The aggregate Bayes risk error, the difference between the detection performance of the decision makers' majority vote and the centralized Bayes optimal risk, is plotted as a function of the signal noise in Fig.~\ref{fig:aggbre}.
We see that this error also first increases and then decreases.

\begin{figure}
	\begin{center}
		\begin{tabular}{c}
			\includegraphics[width=0.47\textwidth]{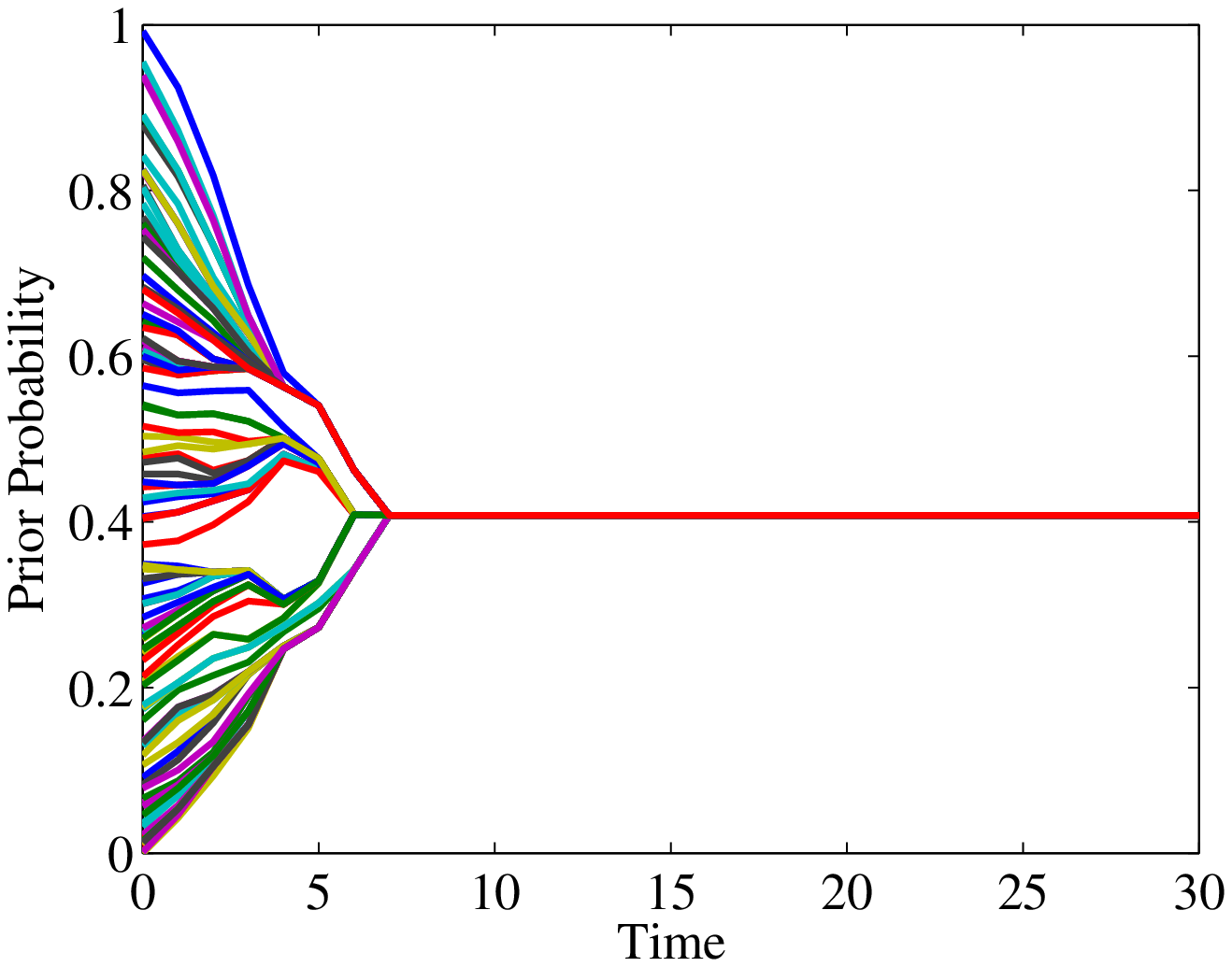} \\
			(a) \\
			\includegraphics[width=0.47\textwidth]{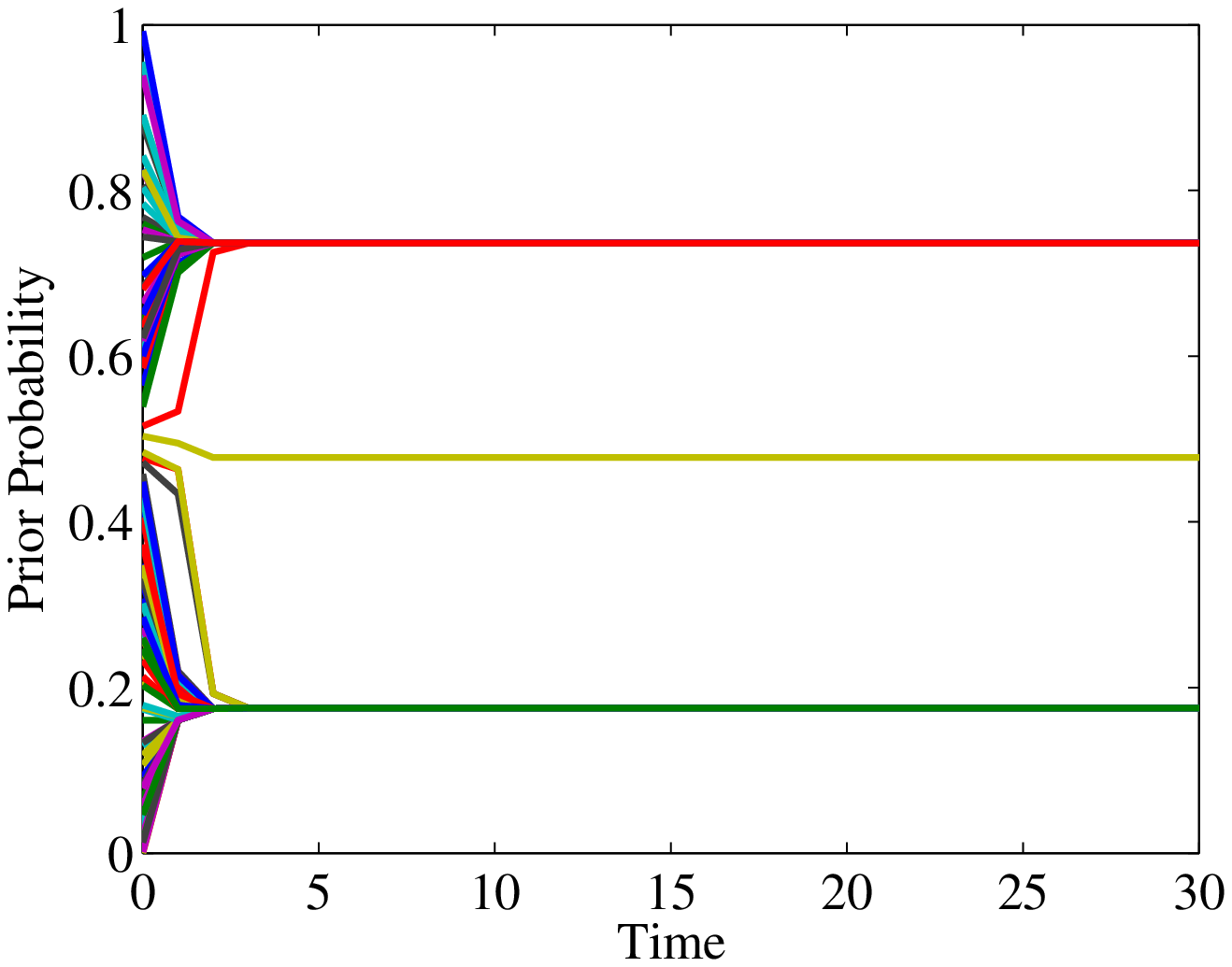} \\
			(b) \\
			\includegraphics[width=0.47\textwidth]{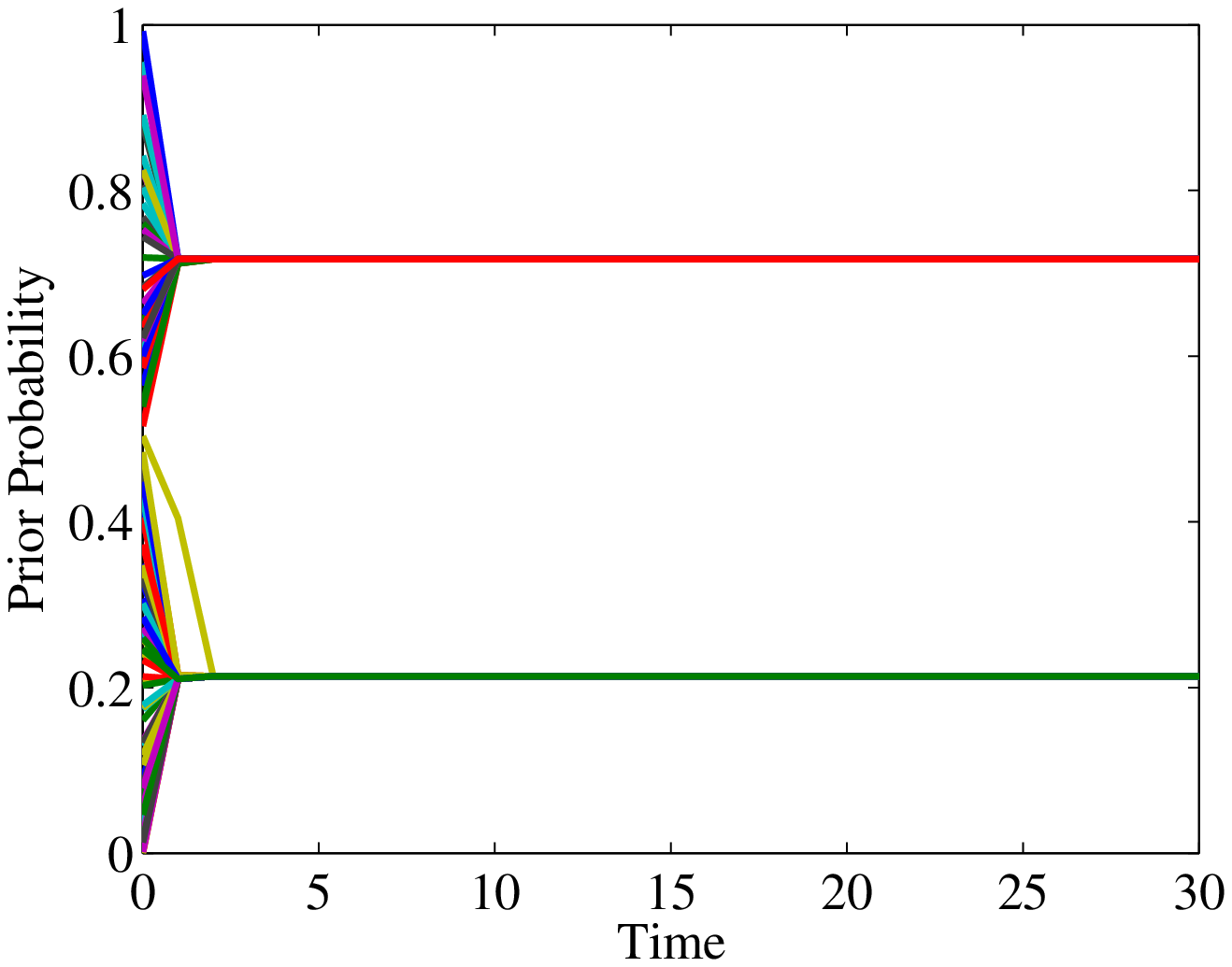} \\
			(c)
		\end{tabular}
	\end{center}
	\caption{Dynamics of 101 people with bounded confidence threshold 0.025 for Bayes risk error divergence Krause-Hegselmann model with univariate Gaussian observations having means 0 and 1, and (a) standard deviation 0.5, (b) standard deviation 4, and (c) standard deviation 32.}
	\label{fig:dynbeta}
\end{figure}

\begin{figure}
	\begin{center}
		\includegraphics[width=0.47\textwidth]{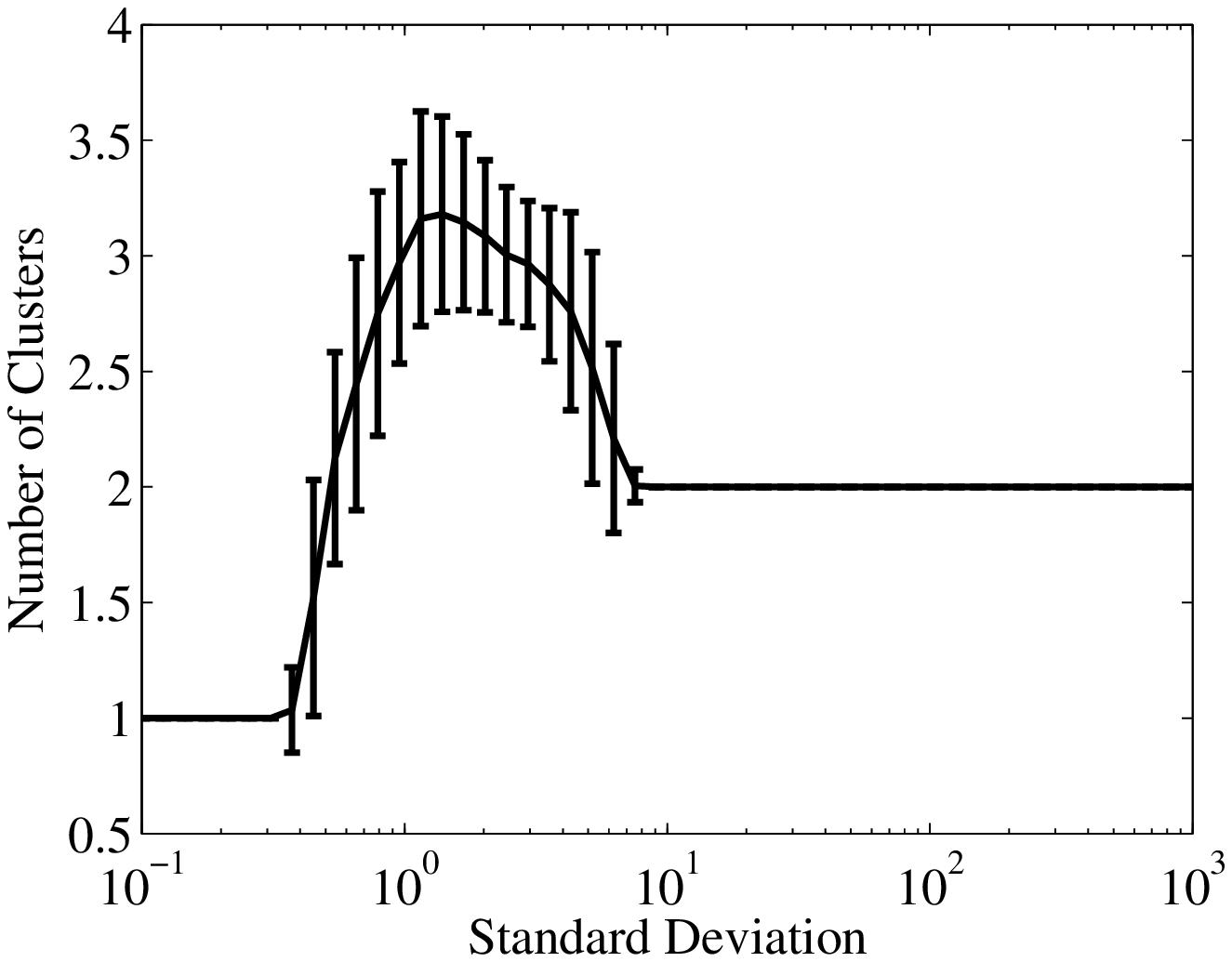}
	\end{center}
	\caption{Converged number of clusters as a function of the noise standard deviation in the second example, averaged over 200 trials.  The error bars indicate the standard deviation of the number of clusters over the trials.}
	\label{fig:numclustersbeta}
\end{figure}

\begin{figure}
	\begin{center}
		\includegraphics[width=0.47\textwidth]{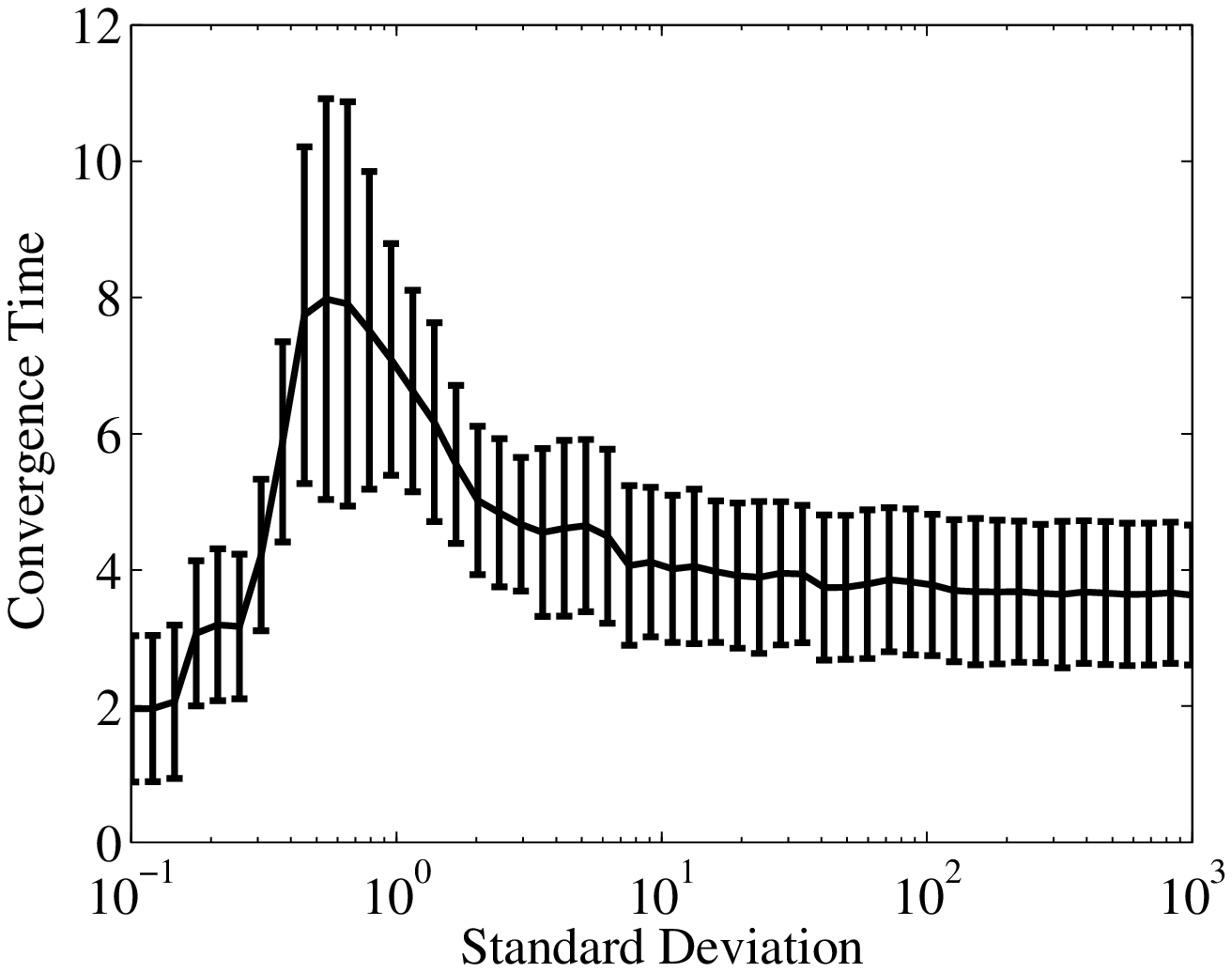}
	\end{center}
	\caption{Convergence time as a function of the noise standard deviation in the second example, averaged over 200 trials.  The error bars indicate the standard deviation of the number of clusters over the trials.}
	\label{fig:numitersbeta}
\end{figure}

\begin{figure}
	\begin{center}
		\includegraphics[width=0.47\textwidth]{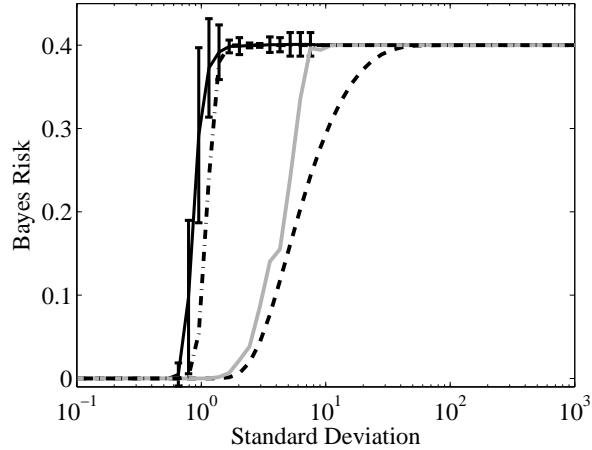}
	\end{center}
	\caption{Bayes risk at convergence of the majority vote of 101 opinion givers as a function of the noise standard deviation in the second example, averaged over 200 trials.  The error bars indicate the standard deviation of the number of clusters over the trials.  The dashed line is the centralized Bayes optimal risk, the gray line is the decentralized Chair-Varshney rule Bayes optimal risk, and the dashed and dotted line is the decentralized majority vote Bayes optimal risk.}
	\label{fig:bayesriskbeta}
\end{figure}

\begin{figure}
	\begin{center}
		\includegraphics[width=0.47\textwidth]{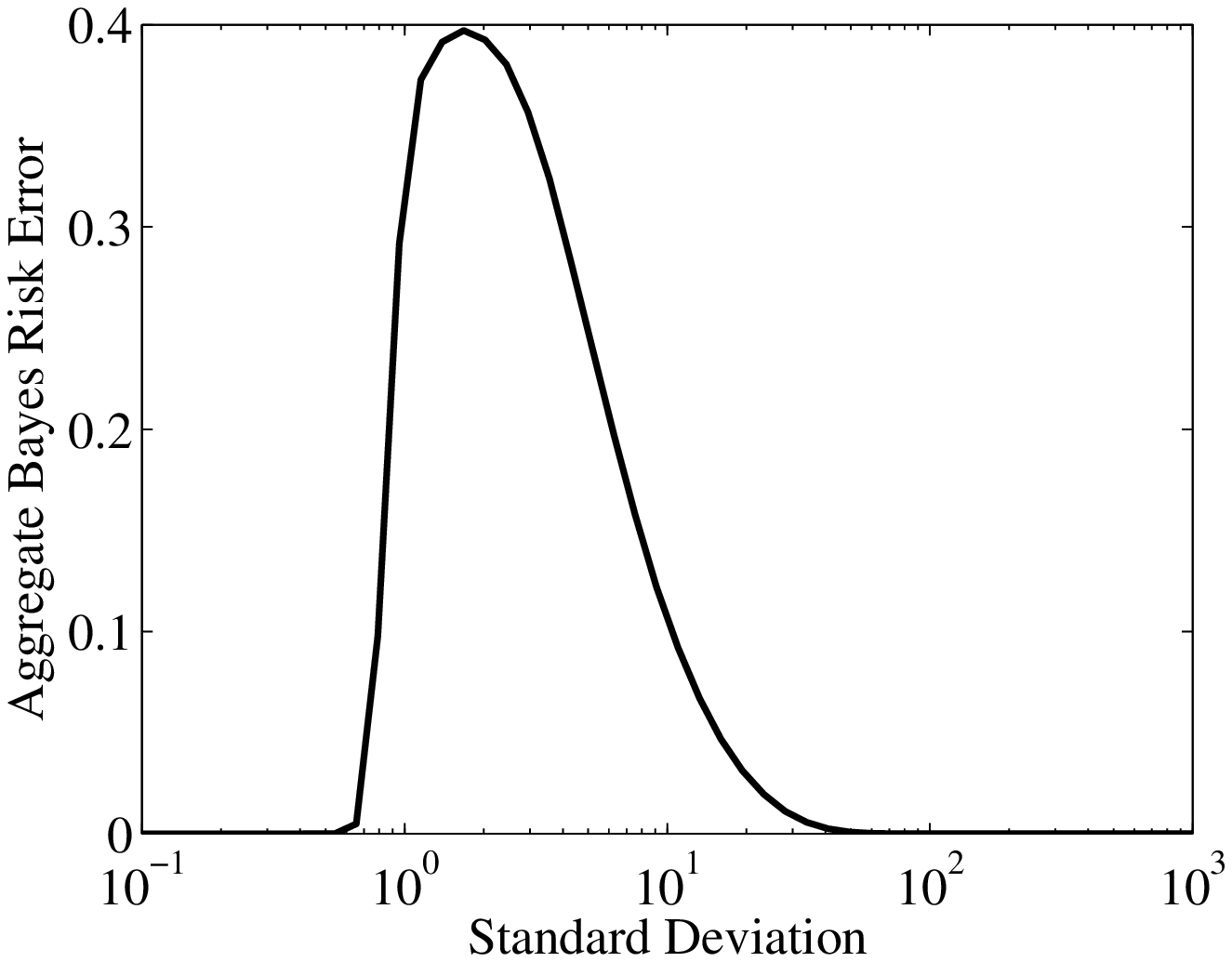}
	\end{center}
	\caption{Aggregate Bayes risk error at convergence of the majority vote of 101 opinion givers as a function of the noise standard deviation in the second example, averaged over 200 trials.}
	\label{fig:aggbrebeta}
\end{figure}

The same plots are presented for the second example with beta-distributed initial beliefs in Fig.~\ref{fig:dynbeta}--Fig.~\ref{fig:aggbrebeta} with $\theta = 0.025$. In this example, the convergence time displays a different character than that seen in the first example, but the behavior of the number of converged clusters is similar.  Also in the Bayes risk plot for this example, Fig.~\ref{fig:bayesriskbeta}, there is one additional performance curve for comparison, the Chair-Varshney optimal fusion rule.  (The Chair-Varshney rule for the first example is precisely the majority vote rule.)

In the plots of the final number of clusters in both examples, we observe an intermediate maximum as a function of signal-to-noise ratio.  Such a phenomenon with the number of converged clusters in bounded confidence opinion dynamics does not appear in any previous work.  The interpretation is that at very low noise, the observation readily indicates the true hypothesis so priors do not matter, and thus everyone converges to an average, maximally ignorant prior.  The intermediate noise regime corresponds to the most contentious issues and opinions because the observation is somewhat informative of the true hypothesis, but not fully.  The choice of prior has a large effect on the opinion, and thus these contentious issues lead to factions.  With very large amounts of noise, no one has any idea what is going on, and everyone is uninformed; again, convergence is to an average, uninformed consensus of priors.  

Although such a phenomenon is known to occur in society, to the best of the author's knowledge, there has been no previously proposed opinion dynamics model that generates more factions when the issue is more contentious.  Moreover, it is in these contentious issues that there is the greatest degradation in the detection performance by the aggregated social network from the various centralized and decentralized Bayes optimal performances.  The convergence time does not seem to have a fixed pattern as a function of signal-to-noise ratio as illustrated by the two examples and other examples not shown here.

\section{Conclusion}
\label{sec:conclusion}

In this paper, we have examined bounded confidence opinion dynamics when the opinions are endowed with a meaning as decisions in a hypothesis testing or signal detection task.  The model that arises in this setting is very similar to the standard Krause-Hegselmann model of opinion dynamics, but with the key difference that the confidence bound is calcuated via Bayes risk error divergence rather than absolute error.  The convergence results and behaviors in this setting follow those of the Krause-Hegselmann model.  

By being endowed with the decision-making context, we are able to examine the behavior of the opinion dynamics as a function of observation noise.  In doing so, we find that the number of converged clusters first increases and then decreases with signal-to-noise ratio in the detection task.  We are also able to examine the decision-making performance of the aggregate group.  In a similar fashion, the aggregate Bayes risk error first increases and then decreases with signal-to-noise ratio.  This phenomenon of the most contentious issues leading to the largest number of clusters or factions is a model prediction that has not and could not appear in any previous work on opinion dynamics.  Interestingly enough, it is the same set of issues that lead to the largest degradation in detection performance evidenced by aggregate Bayes risk error.

In future work, we would like to extend the model to also take social acquaintance edges into account and adapt the theoretical convergence results of \cite{Rabbat2012} to bounded confidence defined according to Bayes risk error divergence.  Under such a social network graph in addition to bounded confidence, we can examine several scenarios of real-world interest by setting the social connections graph, initial decision weights, observation noise, and detection error costs appropriately.  For example, we can examine voting in the United States Congress using its cosponsorship graph as the social network and validating model predictions using actual votes \cite{WangVM2012}.  A second scenario of interest is looking at small-world networks to model impoverished villagers who are forming opinions on vaccination for their children \cite{BanerjeeDGK2010}; the result of vaccination is very uncertain, i.e., has very high observation noise, because of the very long duration until its effect and also has very high missed detection costs.  Another domain that could be examined is elections (especially in homophilic social networks), in which there is evidence that when the observation noise is small, people vote for their ethnicity less, but when observation noise is large, people vote for their ethnicity more \cite{Wantchekon2003}.  

In terms of model extensions, future work could include putting a decision-making twist on any of the bounded confidence opinion dynamics extensions listed in Section~\ref{sec:prelim:previous}, considering people having differing Bayes costs \cite{RhimVG2011}, and considering correlated observations among people rather than statistically independent ones.  Also, if a series of opinions were sought from people, a model in which noise variance decreases over time could be considered.  The theoretical convergence analysis presented herein, and also in the literature for the standard Krause-Hegselmann model has not been able to precisely characterize the number of converged clusters, which is central to the main insight of this paper; theoretical analysis of that characterization is of great importance and should be pursued in future work.  Although it did not show a pattern of behavior, convergence time should similarly be theoretically investigated further \cite{ZhangH2013}.

\section*{Acknowledgment}

The author thanks Michael Rabbat for an initial conversation on opinion dynamics, Lav R.~Varshney and Pramod K.~Varshney for discussions, and Asuman Ozdaglar for encouragement.

\ifCLASSOPTIONcaptionsoff
  \newpage
\fi



\bibliographystyle{IEEEtran}
\bibliography{IEEEabrv,jopinion}
\end{document}